\documentclass{kapproc} 

\RequirePackage{graphicx}%
\RequirePackage{epsf}%
\input{psfig}

\upperandlowercase
\let\footnote\savefootnote
\let\footnotetext\savefootnotetext 
\let\footnoterule\savefootnoterule 
\kluwerbib 

\begin{document}


\articletitle{Age spreads in clusters and associations: the lithium test}


\chaptitlerunninghead{Lithium in the ONC}



 \author{Francesco Palla and Sofia Randich}
 \affil{INAF--Osservatorio Astrofisico di Arcetri, L.go E. Fermi 5, 
 50125 Firenze, Italy}
 \email{palla@arcetri.astro.it,randich@arcetri.astro.it}





 \begin{abstract}
We report the evidence that several low-mass stars ($<$0.4 M$_\odot$) of the
Orion and Upper Scorpius clusters have lithium abundances well below the
interstellar value.  Due to time-dependent depletion, our result implies
stellar ages greater than $\sim$5 Myr, suggesting that star formation has
been proceeding for a long time in these systems.

 \end{abstract}

\section{How long does star formation last in clusters?}
A central debate on star formation (SF) concerns the time scale during which
a molecular cloud complex can sustain the production of new stars. Two
competing views exist: a {\it slow mode} regulated by the quasi-static
evolution of magnetized clouds (e.g., Palla \& Stahler 2000), and a {\it fast
mode} driven by the dissipation of turbulence followed by prompt
gravitational collapse (e.g., Hartmann, this conference). Whether or not SF
can last for an extended period of time ($\sim 10^7$~yr), longer than the
dynamical time (1--2 Myr), is still unclear. Age dating based on isochrones
in the HR diagram of nearby SF regions yields age distributions in the range
1--3 Myr with evidence for a significant population of older
stars ($>$5 Myr). However, given the uncertainties of the isochronal  method,
it is important to find independent ways of gauging stellar ages and age
spreads.

\section{Using Li-depletion to estimate age spreads}
The age dating method based on Li abundances rests on the ability of low-mass
stars to burn their initial Li content during the early pahses of PMS
contraction.  
Lithium depletion time scales vary with stellar mass, being
$\sim$10 Myr for 0.2--0.4 M$_\odot$ stars. The Li-based method has been
successfully applied to relatively young open clusters ($>$30 Myr; e.g.,
Stauffer 2000), but never to SFRs in the assumption that their low-mass
members are too young and too cold for nuclear burning.

We have measured the Li~I 6708 \AA~ line in a sample of $\sim$90 ONC members
with mass 0.4--0.8 M$_\odot$ and isochronal ages greater than $\sim$1 Myr
using FLAMES$+$Giraffe on ESO-VLT2. As shown in Fig. 1, we find a decrease of
the Li-abundance by a factor 5-10 in the coldest (T$\sim$3700~K) and faintest
objects. Comparison with PMS evolutionary models indicates that the observed
Li-depletion corresponds to stellar ages greater than $\sim$5 Myr.

In Upper Scorpius we have determined the EW(Li)-isochronal age relation
using the low-mass stars observed by Preibisch \& Zinnecker (1999):
the oldest stars appear to have a much lower EW(Li) than the bulk of the
younger stars. The most Li-poor stars are also the least massive objects
($\sim$0.2 M$_\odot$), setting an age estimate greater than $\sim$10 Myr.

We conclude that differences in Li-abundance are present among low-mass
members of very young clusters.  Due to the time dependence of Li-depletion,
these variations can be interpreted as a substantial spread in stellar ages.
This finding offers an independent evidence for a slow mode of star
formation. In the near future, we will extend our observations to stars of
lower mass with the goal of detecting the Li-depletion boundary, thus setting
an absolute age for very young clusters.

\begin{figure}[ht]
\centerline{\psfig{file=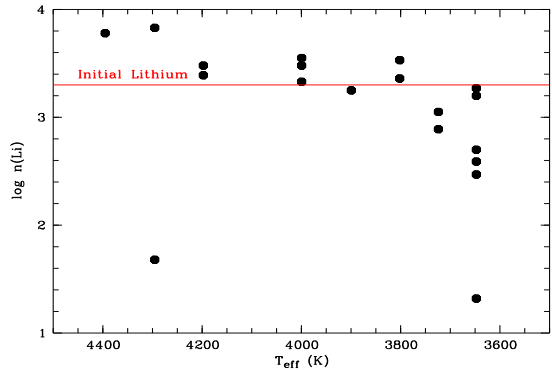,width=3.3in,height=2.1in,angle=-90}}
\caption{Li-abundances in low-mass stars of the Orion Cluster}
\end{figure}

\end{document}